\documentclass[pre,twocolumn,showpacs,superscriptaddress,floatfix]{revtex4}
\usepackage{graphicx}
\usepackage{dcolumn}% Align table columns on decimal point
\usepackage{bm}% bold math

\begin{document}
\title{Anisotropic thermally activated diffusion in percolation systems}
\author{S. Bustingorry}
\affiliation{Consejo Nacional de
Investigaciones Cient\'{\i}ficas y T\'ecnicas, Centro At\'omico Bariloche, (8400) S. C. de Bariloche, Argentina.}
\author{G. L. Insua}
\affiliation{Departamento de F\'{\i}sica, Universidad Nacional del Comahue,
(8300) Neuqu\'{e}n, Argentina.}
\date{\today}

\begin{abstract}

We present a study of static and frequency-dependent diffusion with anisotropic
thermally activated transition rates in a two-dimensional bond percolation
system.
The approach accounts for temperature effects on diffusion
coefficients in disordered anisotropic systems.
Static diffusion shows an Arrhenius behavior for low temperatures with an
activation energy given by the highest energy barrier of the system. From the
frequency-dependent diffusion coefficients we calculate a characteristic
frequency $\omega _{c}\sim 1/t_{c}$, related to the time $t_c$ needed to
overcome a characteristic barrier. We find that $\omega_c$ follows an Arrhenius
behavior with different activation energies in each direction.

\end{abstract}


\maketitle

The study of diffusion on disordered media is an important problem, in view of
its relevance in a wide variety of natural and industrial processes
\cite{haus87,bouc90,sahimiL}.
In the last years the anisotropic generalization of diffusion has attracted 
much attention \cite{parr87,reyes,cacerey,bust00,saad02,huang02},
justified by the diversity of systems in which diffusion takes place.
A few examples of anisotropic systems are porous reservoir rocks 
\cite{sahimiL,saad02,koelman}, epoxy-graphite disk composites \cite{celz94}
and layered semiconducting compounds \cite{gallos}.

Static (long-time) and frequency-dependent conductivity on isotropic disordered
media has been extensively studied with both analytical and numerical methods.
One of the most widely used models for disorder media is the percolation model
\cite{sahimiL,stauffer,grimmet}, due to its simplicity
and interesting properties (characteristic percolation threshold, fractality of
the sample-spanning cluster, etc). More recently, diffusion has also been
studied in anisotropic bond percolation systems \cite{parr87,reyes,bust00}.
However, the interplay between temperature and disorder in these systems has not
been studied yet.
The natural manner to include temperature effects in diffusion problems is through
thermally activated processes, \textit{i.e.}, associating energy barriers to
the transition rates.
Interesting results are found at low temperatures in isotropic systems,
where the competing effects of temperature, energy barriers and
topology become important \cite{horn95,argy95,dyreR}.
Static
diffusion is then described by an Arrhenius law with a characteristic
energy which depends on the percolation threshold of the lattice
\cite{horn95,argy95}, and frequency-dependent diffusion 
becomes universal if properly scaled \cite{dyreR}. The scaled
units include information about different parameters: temperature,
characteristic percolation energy, and a characteristic frequency
that marks the onset of static diffusion.

In this paper we shall focus on the description of a model for anisotropic
diffusion processes, both in the static and frequency-dependent regimes.
In order to emphasize the role of temperature and its relevance for diffusion in
a system with energy disorder, we use here a two-dimensional isotropic bond
percolation system with different energy-dependent intrinsic transition rates
in each direction.
We define the transition rates as $w_{1(2)}$ in the $1(2)$ direction
and choose them according to $\Gamma(w_{1})$ and $\Omega(w_{2})$, the
probability distribution functions (PDFs), which for this anisotropic
model are given by
\begin{eqnarray}
\label{distdes}
\Gamma (w_{1}) =p\delta (w_{1}-w_{1}^{0})
+(1-p)\delta (w_{1}) , \nonumber \\*
\Omega (w_{2}) =p\delta (w_{2}-w_{2}^{0})
+(1-p)\delta (w_{2}).
\end{eqnarray}
It means that the transition rate $w_{1}$
takes the value $w_{1}^{0}$ with occupation probability $p$, and zero otherwise,
and analogously for the $2$ direction.

In order to account for the temperature dependence of diffusion coefficients,
we propose intrinsic transition rates characterized by a
thermally activated process with different activation energies in each
direction.
Therefore we define an \textit{anisotropic thermally activated process},
in which the transition rates in each direction take the form 
\begin{equation}
w_{1(2)}^{0}=\gamma _{0} \exp \left(-\frac{\epsilon _{1(2)}}{kT} \right),
\label{trthermal}
\end{equation}
where $\gamma _{0}$ is the constant jump rate and $k$ is the Boltzmann constant.
The $(1-p)$ fraction of non-conductor components in Eq. (\ref{distdes})
now represents the existence of infinite energy barriers.

We set $\epsilon_{1} > \epsilon_{2}$ and define an anisotropic parameter
$\alpha =\epsilon _{1}/\epsilon _{2}$,
and a mean energy $\epsilon =(\epsilon _{1}+\epsilon _{2})/2$.
This mean energy was kept constant in the present work.
In terms of these parameters we may write
$\epsilon _{1}=2\alpha \epsilon /(\alpha +1)$ and 
$\epsilon _{2}=2\epsilon /(\alpha +1)$.
Then, the relevant parameters of the problem become the
occupation probability $p$, the temperature $T$, and the anisotropic parameter
$\alpha $. In the following, energies and
temperatures are measured in arbitrary units (with $k=1$).

The proposed model is studied both
analytically, by using an anisotropic extension to the
effective medium approximation (EMA), and numerically by means of standard Monte
Carlo (MC) simulations.
We shall briefly describe both methods.

The reader is referred to \cite{reyes} for a complete description of anisotropic
EMA. Here we only summarize the key results. The EMA consists in
averaging the effects of disorder by defining an effective medium with
effective transition rates, which depend on the Laplace variable $u$.
These effective transition rates are self-consistently determined by the
requirement that the difference between the propagator of the impurity and
homogeneous problems should average to zero.
Thus, in anisotropic problems, two effective
transition rates, $w_{1}^{e}(u)$ and
$w_{2}^{e}(u)$ (one for each direction), are introduced. These effective
transition rates are determined by two self-consistent conditions
\cite{parr87,reyes}: 
\begin{eqnarray}
\label{aveani1}
\left\langle {\frac {w_{1}^{e}-w_{1}}
{1+2\left( w_{1}^{e}-w_{1}\right)\left[ G^{1} (u)-G^{0} (u)\right]}}
\right\rangle _{\Gamma (w_{1})}=0, \nonumber \\*
\left\langle {\frac {w_{2}^{e}-w_{2}}
{1+2\left( w_{2}^{e}-w_{2}\right)\left[ G^{2} (u)-G^{0} (u)\right] }}
\right\rangle _{\Omega (w_{2})}=0.
\end{eqnarray}
Here, $G^{1(2)}$ and $G^{0}$ are the non-perturbed anisotropic Green
functions related to the probabilities of moving from the origin to one of its
nearest neighbors in the $1(2)$ direction and the return probability,
respectively.
The impure bond connects two nearest neighbor sites of the lattice whose
transition rates are equal to $w_{1}$ if the impure bond lies in the $1$
direction and $w_{2}$ if the impure bond is in the other direction.

Following the linear response theory \cite{odagaki1} the generalized
frequency-dependent complex
diffusion coefficients $D(\omega)$ in the anisotropic EMA context are given by 
$D_{1(2)}(\omega)=a^{2}\,w_{1(2)}^{e}(u=i \omega)$,
where $a$ is the lattice constant. In the following we will take $a=1$.

In the static case ($\omega =0$) Eqs. (\ref{aveani1}) 
become 
\begin{eqnarray}
\label{emaatau0}
\frac{2}{\pi} \left[ D_{1}^{0} -\gamma _{0}\exp \left( -\frac{\epsilon_{1}}{kT}
\right) \right] \arctan \sqrt{\frac {D_{1}^{0}}{D_{2}^{0} }}&& \nonumber \\*
+p\gamma _{0}\exp \left( -\frac {\epsilon_{1}}{kT}\right)&&=0, \nonumber \\*
\frac{2}{\pi} \left[ D_{2}^{0} -\gamma _{0}\exp \left( -\frac{\epsilon_{2}}{kT}
\right) \right] \arctan \sqrt{\frac {D_{2}^{0}}{D_{1}^{0} }}&& \nonumber \\*
+p\gamma _{0}\exp \left( -\frac {\epsilon_{2}}{kT}\right)&&=0,
\end{eqnarray}
where the transition rates are defined in Eq.
(\ref{trthermal}), $D_{1}^{0}=D_{1}(\omega =0)=w_{1}^{e}(0)$ and
$D_{2}^{0}=D_{2}(\omega =0)=w_{2}^{e}(0)$ represent
zero-frequency (static) diffusion coefficients in each direction.

Monte Carlo data are obtained by performing classical random walk
simulation in square lattices with $300^{2}$ sites. Normal diffusion regime
(long-time  limit) is reached between $10^{3}$ and $10^{6}$ steps, and
the mean square displacement is averaged over $2000$ and $10000$ different
lattices, depending on data fluctuations. Static diffusion
coefficients are obtained via the Einstein's relations, $\left\langle
R_{1(2)}^{2}\right\rangle =2D_{1(2)}^{0}t$.
As was shown earlier \cite{dyre94}, standard MC simulations fail to
describe particle diffusion for very low temperatures.
In this limit the method is inefficient and other simulation methods are
necessary. Here the standard MC method was used and simulations were
performed for temperatures above $T=0.05$.

We present now the results for the long-time diffusion properties.
Firstly, in the isotropic case $\alpha =1$, the transition rates are given by
$\gamma _{0}\exp \left( -\frac{\epsilon}{kT}\right) $
and, as expected, only one diffusion coefficient is obtained, viz.,
$D_{iso}^{0}=D_{1}^{0}=D_{2}^{0}=\gamma _{0}\exp \left( -\frac{\epsilon }{kT}\right) (2p-1)$.
Secondly, for all values of $\alpha$, in the $T\rightarrow \infty $ limit
($kT\gg \epsilon $) the model
again reduces to the isotropic bond percolation problem with transition
rates $\gamma _{0}$ [Eq. (\ref{trthermal})].
In this case, the isotropic diffusion coefficient is given by
$D_{iso}^{0} =\gamma _{0}(2p-1)$.
Finally, for values of $\alpha \neq 1$ and in the low temperature limit
$D_{1}^{0}$ and $D_{2}^{0}$ were calculated by numerically solving the set of 
Eqs. (\ref {emaatau0}), and results are compared with MC simulations.

\begin{figure}[!tbp]
\includegraphics[width=8.5cm,clip=true]{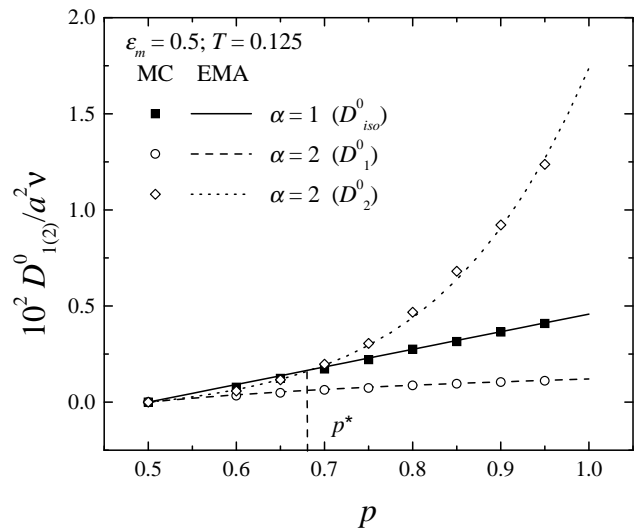}
\caption{\label{fig1}
Dependence of the anisotropic static diffusion coefficients on
the occupation probability $p$ for $T=0.125$.
The characteristic value $p^{*}$ is also indicated.
%Diffusion coefficients are presented in
%units of squared lattice parameter times jump frequency.
Energy and temperature are given in arbitrary units with the Boltzmann constant
$k=1$.}
\end{figure}

Let us consider the variation of $D_{1}^{0}$ and $D_{2}^{0}$ with the
occupation probability $p$, for fixed values of $T$ in anisotropic conditions.
The calculation was performed for $\epsilon = 0.5$ and $\gamma_{0}=0.25 \nu$,
where $\nu$ is a characteristic jump frequency.
Figure \ref{fig1} shows $D_{1(2)}^{0}$ as a function of $p$ for $T=0.125$.
Symbols represent MC simulations and lines correspond to the EMA numerical
solution of Eqs. (\ref{emaatau0}).
The isotropic $\alpha =1$ case is included for comparison.
For $\alpha>1$, we obtain $D_{1}^{0}<D_{2}^{0}$, which is in agreement with the
fact that lower energy barriers imply higher diffusion coefficients.
For $p\leq p_{c}=0.5$, the isotropic bond percolation
threshold \cite{stauffer}, we find that $D_{1}^{0}=D_{2}^{0}=0$ as expected.
In the low $p$ region the two anisotropic diffusion coefficients are lower than
the isotropic one ($p\leq p^{*}\simeq 0.68$). In this region
the conducting cluster is poorly connected and the diffusion in the low
energy direction is highly affected by the existence and height of high
energy barriers. As $p$ is increased, more energy barriers appear and the
diffusion in the low energy barrier direction is less sensitive to high
energy barriers. This kind of behavior is also found for other values of $T$,
with increasing values of $p^{*}$ as $T$ decreases.

\begin{figure}[!tbp]
\includegraphics[width=8.5cm,clip=true]{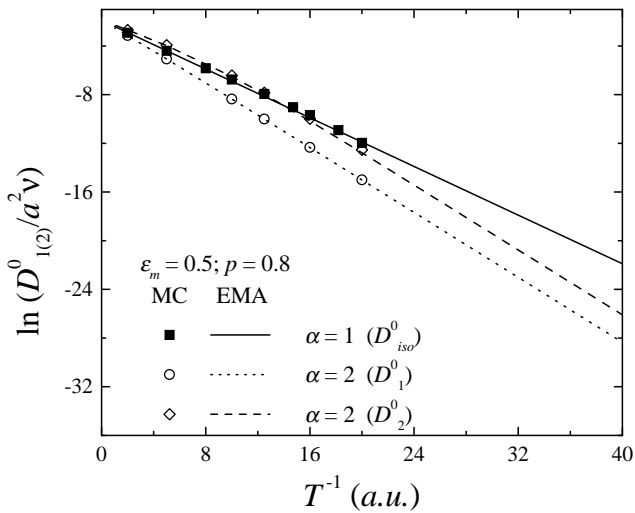}
\caption{\label{fig2}
Arrhenius plot for the anisotropic static diffusion coefficients
at low temperatures for $p=0.8$.
%Diffusion coefficients are presented in units
%of $a^2 \nu$ and temperature are given in arbitrary units.
Energy and temperature are given in arbitrary units with the Boltzmann constant
$k=1$.}
\end{figure}

Next we turn to the behavior of $D_{1}^{0}$ and $D_{2}^{0}$ at low temperatures
for fixed values of $p$.
We do not present here results for $p=p_{c}$ (or near $p_{c}$), the percolation
threshold, where the
substrate has fractal properties and diffusion becomes anomalous
\cite{haus87,bouc90}.
In Fig. \ref{fig2} we show the temperature dependence of the static
diffusion coefficients in an Arrhenius plot for $p=0.8$ and $\alpha =1$ and $2$.
From the EMA curves (lines) the slope of the linear region is numerically
calculated.
For isotropic media, the EMA predicts a slope equal to $-0.5=-\epsilon$.
For $\alpha =2$, we find that $D_{1}^{0}(T)$ and $D_{2}^{0}(T)$ reach an
asymptotic Arrhenius behavior with the same activation energy
$\epsilon _{1}$, corresponding to the highest energy of this
anisotropic model.
This is not a striking result considering that long-time diffusion at low
temperatures is strongly governed by the highest energy barrier, as
the particle spends a lot of time trying to overcome them while
performing long-distance trajectories.
The low temperature behavior of static diffusion coefficients depends only on
the highest characteristic energy value of this anisotropic model.

\begin{figure}[!tpb]
\includegraphics[width=8cm,clip=true]{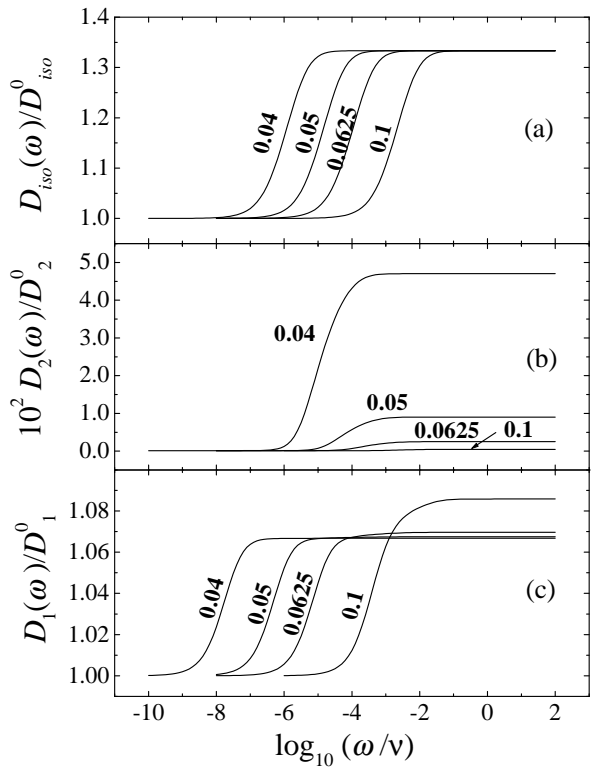}
\caption{\label{fig3}
Scaled frequency-dependent diffusion coefficients for isotropic
(a) and anisotropic $\alpha=2$ [(b) and (c)] cases. Each curve is
labeled with its corresponding temperature (in arbitrary units).
%Frequency is presented in units of $\nu$.
}
\end{figure}

The frequency behavior of diffusion coefficients was studied by numerically
solving the set of coupled Eqs. (\ref{aveani1}),
using the corresponding expressions for the anisotropic Green functions
\cite{reyes}, and the temperature dependence introduced in Eq. (\ref{trthermal}).
We present here results for the real part of the frequency-dependent complex
diffusion coefficients.
In Fig. \ref{fig3} we show the scaled values 
$D_{1(2)}(\omega) / D_{1(2)}^{0}$
for $p=0.8$, $\epsilon =0.5$, and temperature values between $0.1$
and $0.04$. Three characteristic regimes can be distinguished:
(\textit{i}) the long-time limit for $\omega \rightarrow 0$, (\textit{ii}) a
power law behavior characteristic of intermediate frequencies, $\sigma \sim
\omega ^{s}$, with $s\leq 1$ and, (\textit{iii}) the high-frequency regime,
$\omega \rightarrow \infty $, where the diffusion coefficients approach to a
constant value.
This kind of dielectric response has been observed before in a broad class
of ionic and electronic systems, in isotropic and anisotropic media
\cite{jhan96,angeR,dyreR}.

In the isotropic case we find that while decreasing the temperature the scaled
value $D_{iso}(\omega)/D_{iso}^{0}$ reaches a temperature-independent value
for $\omega \rightarrow \infty$ [Fig. \ref{fig3} (a)].
In the anisotropic case, while $D_{1}(\omega)/D_{1}^{0}$ saturates for
$\omega \rightarrow \infty$ to a temperature-independent value at low
temperatures [Fig.\ref{fig3} (c)],
$D_{2}(\omega)/D_{2}^{0}$ saturates to a temperature-dependent value that
follows an Arrhenius law with an activation energy equal
to $\epsilon_{1}-\epsilon_{2}$, even at low temperatures [Fig.\ref{fig3} (b)].
This $\omega \rightarrow \infty$ behavior can be interpreted from the evolution
of the particle diffusion for $t \rightarrow 0$.
For $\alpha =1$ we find that
$D_{iso}(\omega \rightarrow \infty) = p \gamma_{0} \exp \left(-
\frac {\epsilon}{kT} \right)$.
This means that the first step of the particle is given by the probability $p$
of finding a given bond times the intrinsic transition rate of that bond.
From this considerations and using the isotropic result for $D_{iso}^{0}(T)$ we
find that $D_{iso}(\omega \rightarrow \infty )/D_{iso}^{0}=p/(2p-1)=4/3$ for
$p=0.8$, as shown in Fig. \ref{fig3} (a).
In anisotropic conditions, we expect the $t \rightarrow 0$
($\omega \rightarrow \infty$) evolution to be given
by $p \gamma_{0} \exp \left( -\frac{\epsilon_{1}}{kT} \right)$ in the $1$
direction and by $p \gamma_{0} \exp \left( -\frac{\epsilon_{2}}{kT} \right)$
in the $2$ direction.
We showed previously that
$D_{1}^{0}\sim \exp \left( -\frac{\epsilon _{1}}{kT} \right) $,
and $D_{2}^{0}\sim \exp \left( -\frac{\epsilon _{1}}{kT}\right) $,
thus $D_{1}(\omega \rightarrow \infty )/D_{1}^{0}$ is expected to be constant
at low temperatures, and $D_{2}(\omega \rightarrow \infty )/D_{2}^{0}\sim 
\exp \left( \frac{\epsilon _{1}-\epsilon _{2}}{kT}\right)$,
which account for the calculated EMA results for $\omega \rightarrow \infty$
in Fig. \ref{fig3} (b) and (c).

\begin{figure}[!tpb]
\includegraphics[width=8.5cm,clip=true]{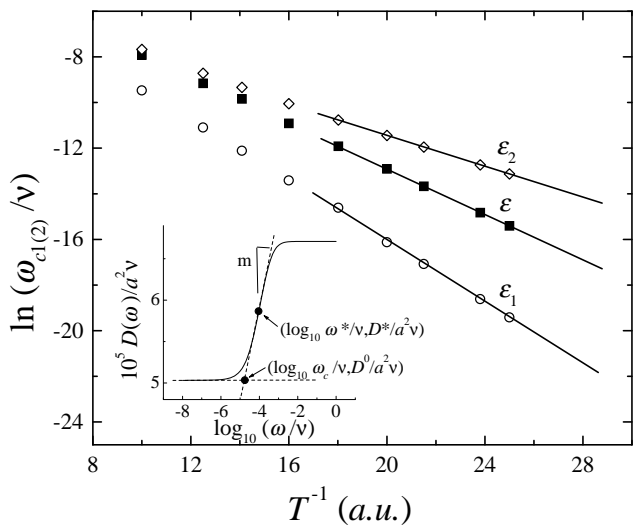}
\caption{\label{fig4}
Temperature dependence of the characteristic frequencies 
$\omega _{c1(2)}$.
The isotropic case $\omega_{c,iso}$ (full squared) and the anisotropic 
$\alpha=2$
cases $\omega_{c1}$ (open circles) and $\omega_{c2}$ (open diamond) are shown.
Lines are guides to the eye and are labeled by its corresponding activation
energy.
The inset shows the determination of the value $\omega_c$ from a typical curve
for the frequency-dependent diffusion coefficient.
The value $m$ represent the slope of the tangent curve through the inflection
point $(\log_{10} \omega ^{*}/\nu,D^{*}/a^2\nu)$ (see text).
%Frequency is presented in units of $\nu$.
}
\end{figure}

For intermediate frequencies a characteristic frequency
$\omega_{c}\sim 1/t_{c}$ can be defined, related to the time $t_{c}$ 
needed to overcome a characteristic energy barrier.
The inset of Fig. \ref{fig4} shows a typical
frequency-dependent $D(\omega)$ plot (as those in Fig. \ref{fig3}).
The $\omega _{c}$ parameter is calculated from our
anisotropic EMA data as follows.
Let the inflection point be $(\log_{10} \omega ^{*}/\nu,D^{*}/a^2\nu)$, and
$m$ the slope of the tangent curve passing through the inflection point.
The $\omega _{c1(2)}$ parameter is given by the intersection of the
slope $m$ through the inflection point and the line corresponding
to the long-time diffusion coefficient $D_{1(2)}^{0}$ (Fig. \ref{fig4}, inset).
We include the subscript $1(2)$ as $\omega _{c}$ may be different
for each direction.
In Fig. \ref{fig4} we present the temperature dependence of the characteristic
frequencies $\omega_{c1(2)}$ in an Arrhenius plot for $\alpha =1$ and $2$.
All cases reach a linear
behavior for low temperature values, but different slopes are found in each
case, corresponding to the characteristic energy in each direction.
This indicates, as expected, that different times
$t_{c1(2)} \sim \exp \left( -\frac{\epsilon_{1(2)}}{kT} \right)$ are needed in
each direction to overcome its characteristic energy barriers.

In summary, in the present paper we described long-time and frequency-dependent
diffusion in anisotropic thermally activated processes in a two-dimensional
bond percolation lattice.
The present model has different activation energies in each
direction of the square lattice.
Long-time diffusion coefficients follow an Arrhenius law at low temperatures
with the highest energy barrier being the activation energy for diffusion in
both directions of the lattice.
From the frequency-dependent diffusion coefficients, we define a characteristic
frequency $\omega_{c1(2)}$.
We remark that $\omega_{c}$ is not marking the onset of long-time diffusion,
as in isotropic problems with continuous PDFs \cite{dyreR}, but it
is just the frequency associated to overcome the characteristic energy of
\textit{each} direction.
The existence of two different activation energies for $\omega_{c1(2)}$ and
only one for $D_{1(2)}^{0}$ is a consequence of the strength of
the percolation-type disorder, in contrast with results from
isotropic continuous PDFs, where the characteristic frequency and the static
diffusion coefficient follow Arrhenius laws \textit{with the same}
activation energy \cite{horn95,argy95,dyreR}.

S. B. thanks a fellowship from Universidad Nacional del Comahue at the early 
stages of this work. This work was partially suported by Universidad Nacional
del Comahue and CONICET.

\bibliography{Bustin}

\end{document}